\documentstyle{article}

\font\tenrm=cmr10
\font\tenit=cmti10
\font\elevenbf=cmbx10 scaled\magstep 1
\font\elevenrm=cmr10 scaled\magstep 1
\font\elevenit=cmti10 scaled\magstep 1

\font\ninerm=cmr9

\renewenvironment{thebibliography}[1]
 { \elevenrm
   \begin{list}{\arabic{enumi}.}
    {\usecounter{enumi} \setlength{\parsep}{0pt}
     \setlength{\itemsep}{3pt} \settowidth{\labelwidth}{#1.}
     \sloppy
    }}{\end{list}}

\begin{document}
\vspace{-2cm}
\rightline{{ CERN--TH.6922/93}}
\rightline{{ IEM--FT--73/93}}
\rightline{{ FTUAM 23/93}}
\begin{center}
\vglue 0.6cm
 {{\elevenbf        \vglue 10pt

SUSY SOFT BREAKING TERMS FROM STRING SCENARIOS
\footnote{
\ninerm\baselineskip=11pt Talk given at the Int. Workshop on
Supersymmetry and Unification of fundamental Interactions (SUSY-93),
Boston, March 29 -- April 2, 1993
}
\\}
\vglue 1.0cm
{\tenrm Beatriz de CARLOS \\}
\baselineskip=13pt
{\tenit Instituto de Estructura de la Materia (CSIC),\\}
\baselineskip=12pt
{\tenit Serrano 123, 28006--Madrid, Spain\\}}

\vglue 0.3cm
{\tenrm J. Alberto CASAS\footnote{
\ninerm\baselineskip=11pt On leave from Instituto de Estructura
de la Materia (CSIC),
Serrano 123, 28006--Madrid, Spain
}\\}
{\tenit CERN, CH--1211 Geneva 23, Switzerland\\}

\vglue 0.3cm
{\tenrm and\\}
\vglue 0.3cm
{\tenrm Carlos MU\~NOZ\\}
{\tenit Dept. de F\'{\i}sica Te\'orica C-XI, \\}
\baselineskip=12pt
{\tenit Univ. Aut\'onoma de Madrid, E-28049 Madrid, Spain\\}
\vglue 0.8cm
{\tenrm ABSTRACT}

\end{center}

\vglue 0.3cm
{\rightskip=3pc
 \leftskip=3pc
 \tenrm\baselineskip=12pt
 \noindent
The general SUSY soft breaking terms
for a large class of phenomenologically relevant string
scenarios (symmetric orbifolds) are given.
They show a certain lack of universality, but not dangerous
for flavor changing neutral currents.
To get more quantitative results a specific SUSY breaking
mechanism has to be considered, namely
gaugino condensation in the hidden sector. Then, it turns out
that squark and slepton masses
tend to be much larger than scalar masses
($m_{\phi}\stackrel{>}{{}_\sim}10M_a$), which probably is a quite general
fact. Experimental
bounds and the requirement of a successful electroweak breaking
without fine tuning impose further restrictions on the soft breaking
terms. As a consequence
the gluino and chargino masses
should
be quite close to their present experimental limits, whereas
squark and slepton masses
should be much higher
($\stackrel{>}{{}_\sim}\ 1$ TeV).}

\vglue 2cm
\begin{flushleft}
{CERN--TH.6922/93} \\
{IEM--FT--73/93} \\
{FTUAM 23/93} \\
{June 1993}
\end{flushleft}
%

\newpage
\pagestyle{plain}
\pagenumbering{arabic}

{\elevenbf\noindent 1. Introduction}
\vglue 0.4cm
\baselineskip=14pt
\elevenrm

In the last three years there has been substantial progress in our
understanding of supersymmetry (SUSY) breaking in string
theories${}^{1-7}$. In particular, it has been learned that gaugino
condensation
effects in the hidden sector are able to break SUSY at a hierarchically
small scale, at the same time as the dilaton $S$ and the moduli $T_i$
acquire reasonable vacuum expectation values (VEVs).
(The VEV of $S$ is related to the value of the gauge coupling constant,
while those of $T_i$ define the size and shape of the compactified space.)
This has been realized in the context of symmetric orbifold constructions.
The modular invariance in the target space${}^{3-5}$ as well as the presence
of matter in the hidden sector play an important role in this
process${}^{2,6,7}$. On the other hand, the SUSY soft breaking terms in
the observable sector are the phenomenological signature of any SUSY
breaking mechanism, since they are responsible of the supersymmetric
particle masses and the electroweak breaking process.
Therefore, it is essential to completely calculate the soft breaking terms
of these scenarios and study their phenomenological viability.
The aim of this talk is to present some recent developments on this
line of research${}^{8}$.

In section 2 we give the general form of the soft SUSY breaking terms
for symmetric orbifold constructions. Interestingly enough, they
show a certain lack of universality, unlike the usual
assumptions of the minimal supersymmetric standard model.
In order to get more quantitative results a specific SUSY breaking
mechanism has to be considered. This is done in section 3
assuming the above mentioned gaugino condensation in the hidden
sector\footnote{A recent different analysis can be found in refs.[9,10]}
(the only mechanism so far analyzed capable to generate
a hierarchical SUSY breakdown in string constructions). Then the
lack of universality of the soft breaking terms can be conveniently
quantified. Furthermore, we observe a quite general fact: gaugino
masses tend to be much smaller than scalar masses.
In section 4 we investigate the phenomenological viability
of the soft breaking terms.
There are two types of tests that the soft breaking terms should pass.
First, they have to be consistent with the experimental (lower) bounds
on gaugino masses, squark masses, etc. Second, they should be small enough
not to spoil the SUSY solution to the gauge hierarchy problem, guaranteeing
a successful electroweak
breaking. The latter
is a naturalness requirement. We will show that the stringy scenarios
considered are consistent with both tests, and that
this imposes very interesting constraints on the parameters defining
the SUSY breakdown. Finally, in section 5 we present our conclusions.
\vglue 0.6cm
{\elevenbf\noindent 2. General characteristics}
\vglue 0.4cm
Following the standard notation we define the soft breaking terms
in the observable sector by
\begin{eqnarray}
{\cal L} = {\cal L}_{{\mathrm {SUSY}}}+{\cal L}_{{\mathrm {soft}}}\;\;.
\label{Lobs}
\end{eqnarray}
Here ${\cal L}_{{\mathrm {SUSY}}}$ is the supersymmetric Lagrangian derived
from the observable superpotential $W_{{\mathrm {obs}}}$, which includes
the usual Yukawa terms $W_Y$ and a mass coupling $\mu H_1H_2$
between the two Higgs doublets $H_1$, $H_2$. Assuming canonically
normalized fields, ${\cal L}_{{\mathrm soft}}$ is given by
\begin{eqnarray}
{\cal L}_{{\mathrm{soft}}} = -\sum_\alpha m_{\alpha}^2 |\phi_\alpha|^2
-\frac{1}{2} \sum_{a=1}^3 M_a \bar\lambda_a\lambda_a
-\left(Am_{3/2}W_Y + Bm_{3/2}\mu H_1H_2
 \ +\ \mathrm{h.c.}\right)
\label{Lsoft}
\end{eqnarray}
where $m_{3/2}$ is the gravitino mass, $\phi_\alpha$
represent the scalar components of the
supersymmetric particles, and $\lambda_a$ are the $U(1)_Y,
SU(2),SU(3)$ gauginos.

The characteristics of soft SUSY breaking terms in the observable
sector are, to a great extent, determined by the type of $N=1$ SUGRA
theory which appears in four dimensions.
As it is known, this is
characterized by the gauge kinetic function $f$, the K\"{a}hler
potential $K$ and the superpotential $W$. It is also customary
to define ${\cal G}=K+\log|W|^2$. These
functions are determined, in principle, in a given
compactification scheme, although in practice they are sufficiently
well known only for orbifold compactification schemes, which
on the other hand have proved to possess very attractive features
from the phenomenological point of view${}^{11}$. More precisely,
in the general case when the gauge group contains several factors $G =
\prod_{a} G_{a}$, the exact gauge kinetic functions in string
perturbation theory, up to small field--independent contributions,
are${}^{12,13}$
\begin{eqnarray}
f^a = k^aS + \frac{1}{4\pi^2}\sum_{i=1}^3 (\frac{1}{2}b_i^{'a} - k^a
\delta_{i}^{GS}) \log(\eta(T_{i}))^{2}\;\;,
\label{fw}
\end{eqnarray}
with
\begin{eqnarray}
b_{i}^{'a} = C(G^a)-\sum_\Phi T(R_\Phi^a)(1+2n_\Phi^i)\;\;,
\label{bp}
\end{eqnarray}
where the meaning of the various quantities appearing in
Eqs.(\ref{fw},\ref{bp}) is the following:
$k^{a}$ is the Kac--Moody level of the $G^{a}$ group ($k^a=1$ is
a very common possibility), $S$ is the dilaton field,
$T_{i}$ ($i = 1, 2, 3$) are untwisted moduli whose real parts
give the radii of the three compact complex dimensions of the
orbifold (Re $T_{i}\propto R_{i}^{2}, i = 1, 2, 3$),
$\delta_{i}^{GS}$ are 1--loop contributions coming from the
Green--Schwarz mechanism, which have been
determined for the simplest (2,2) $Z_{N}$ orbifolds${}^{13}$;
$\eta(T_{i})$ is the Dedekind function, $\Phi$ labels the matter fields
transforming as $R^a_\Phi$ representations under $G^a$, and $n_\Phi^i$ are the
corresponding modular weights. $C(G)$ denotes the Casimir operator
in the adjoint representation of $G$ and $T(R)$ is defined by
Tr$(T^iT^j)=T(R)\delta^{ij}$.
The K\"{a}hler potential $K$ is given by${}^{14,13}$
\begin{eqnarray}
K = -\log Y - \sum_{i=1}^3 \log(T_{i} + \bar{T_{i}})
+\ \sum_\alpha \prod_i (T_i+\bar T_i)^{n_\alpha^i} |\phi_\alpha|^2\
+\ O(|\phi_\alpha|^4)
\;\;,
\label{K}
\end{eqnarray}
where $\phi_\alpha$ are the matter fields and
\begin{eqnarray}
Y\equiv S + \bar S + \frac{1}{4\pi^2}\sum_{i=1}^3
\delta_{i}^{GS}\log(T_{i} + \bar{T_{i}}))
\;\;,
\label{K01}
\end{eqnarray}
which can be considered${}^{13}$
as the redefined gauge coupling constant (up to threshold
corrections) at the unifying string scale
($Y=2g_{\mathrm{str}}^{-2}$).
Finally, the perturbative superpotential $W^{p}$ at the
renormalizable level has the form
\begin{eqnarray}
W^{p} = h_{IJK} \Phi^{(u)}_{I}\Phi^{(u)}_{J}\Phi^{(u)}_{K} +
h^{'}_{IJK}(T_{i}) \Phi^{(t)}_{I}\Phi^{(t)}_{J}\Phi^{(t)}_{K} +
h^{''}_{IJK} \Phi^{(u)}_{I}\Phi^{(t)}_{J}\Phi^{(t)}_{K}
\;\;,
\label{Wpert}
\end{eqnarray}
where $\Phi^{(u)}_{I}$ $(\Phi^{(t)}_{I})$ are untwisted (twisted) charged
matter fields.
The value of $h_{IJK}$, $h^{''}_{IJK}$ for the allowed couplings
is simply a constant, while $h^{'}_{IJK}(T_{i})$ are complicated but
known${}^{15}$ functions of $T_{i}$. Besides $W^{p}$, there is a
non--perturbative piece, $W^{np}$, usually triggered by gaugino
condensation effects in the hidden sector, which is crucial to break
SUSY. In the following
we will assume $\langle W^{p}\rangle =0$, $\langle W^{np}\rangle\neq 0$,
as it happens in all
SUSY breaking scenarios so far analysed. $W^{np}$ depends on the
$S$ and $T$ fields and, sometimes, on certain matter fields $A$, which
are singlet under the relevant gauge groups${}^{6,7}$. As has been
shown${}^{7}$,
the condition $\partial W/\partial A=0$
is the correct one to integrate out these fields. In consequence, we can
use
\begin{eqnarray}
W^{np} = W^{np}(S,T_i)
\label{Wnp}
\end{eqnarray}
without any loss of generality. Expressions (\ref{fw}--\ref{Wnp}) are to
be understood at the string scale${}^{12}$
$M_{Str}=0.527\times g \times 10^{18}
\ $GeV, where $g\simeq 1/\sqrt{2}$ is the corresponding value of
the gauge coupling constant.

The next step is to calculate, once SUSY breaking is assumed, the
form of the soft breaking terms based on Eqs.(\ref{fw}--\ref{Wnp}).
This can be done by comparing the form of the corresponding ${\cal
L_{SUGRA}}$${}^{16}$ with (\ref{Lobs},\ref{Lsoft}).
We give below the general expressions for all the soft breaking
terms${}^{8}$.
(Previous work on the subject can be found in refs.[3,17,5].)
These are given in the usual overall modulus simplification $T=T_1=T_2=T_3$,
the subindex $\phi$ ($\phi=S,T$) denotes $\partial /\partial\phi$ and
$\delta^{GS}=\sum_i \delta^{GS}_i$. Gaugino and scalar masses are obtained
for the canonically normalized fields.

\vspace{0.2cm}
\noindent {\elevenit Gaugino masses:}
\begin{eqnarray}
M_a(Q) & = & 2\pi\alpha_{a}(Q) m_{3/2} \left\{ k^{a}Y^{2}
\left(-\frac{1}{Y}+\frac{\bar{W}_{\bar S}}{\bar W}\right)
+ \frac{Y(T+\bar T)^2}{3Y+\frac{\delta^{GS}}{4 \pi^2}}
\frac{\hat{G}_{2}(T)}{\pi}
\right.
 \nonumber \\
 & \times  & \left.
\left(\sum_{i} \frac{b_i^{'a}}{16 \pi^2}-
k^a\frac{\delta^{GS}}{8 \pi^2}\right)
\left[ \frac{1}{T+\bar T}
\left(3+\frac{\delta^{GS}}{4 \pi^2}
\frac{\bar{W}_{\bar S}}{\bar W}\right)
-\frac{\bar W_{\bar T}}{\bar W}\right]\right\}
\;\;,
\label{M12}
\end{eqnarray}
where $m_{3/2}=e^{K/2}|W|$, $Q$ is the renormalization group scale and
$\hat G_2(T)=-(\frac{2\pi}{T+\bar T}+4\pi\eta^{-1}
\frac{\partial\eta}{\partial T})$.

\vspace{0.2cm}
\noindent {\elevenit Scalar masses:}
\begin{eqnarray}
m^2_{{\phi}_{\alpha}} = V_o\ +\ m_{3/2}^{2}\left[
1+\frac{n_{\alpha}Y^{2}}{(3Y+\frac{\delta^{GS}}{4\pi^2})^{2}}
\left| 3+\frac{\delta^{GS}}{4\pi^2}\frac{W_S}{W}-(T+\bar T)\frac{W_T}{W}
\right|^2 \right]
\;\;,
\label{mQ}
\end{eqnarray}
where $V_o=\langle V\rangle$ is the value
of the cosmological constant, which we will assume to vanish through
the paper.

\vspace{0.2cm}
\noindent {\elevenit Trilinear scalar terms:}

A term $h\Phi_1\Phi_2\Phi_3$ in the  superpotential induces a trilinear
scalar term of the form
\begin{eqnarray}
{\cal L}_{tril} & = & -m_{3/2}\ A\ \hat h\phi_1\phi_2\phi_3
+ {\mathrm{h.c.}}
\nonumber \\
& = & -m_{3/2} \left\{
\left(1-Y\frac{\bar{W}_{\bar S}}{\bar{W}}\right)
 + \left(3+
\frac{\delta^{GS}}{4\pi^{2}}\frac{\bar W_{\bar S}}{\bar W}-(T+\bar{T})
\frac{\bar W_{\bar T}}{\bar W} \right)\right.
\nonumber \\
& \times & \left. \frac{Y}{3Y+\frac{\delta^{GS}}{4\pi^{2}}}\left[
(3+\sum_{\alpha=1}^3 n_{\alpha}) -
 (T+\bar{T})
\frac{h_{T}}{h}\right]
  \right\} \hat h \phi_{1}\phi_{2}\phi_{3} + {\mathrm {h.c.}}
\;\;,
\label{Ltri}
\end{eqnarray}
where $\phi_\alpha$  are the (properly normalized) scalar
components of the respective superfields, $n_\alpha$
are the corresponding modular weights, and  $\hat h =
e^{K/2}\prod_{\alpha=1}^3 (K_{\alpha \bar \alpha})^{-1/2}h$ is the
effective Yukawa coupling between the physical fields.
Notice that for the untwisted case
[$n_\alpha=-1$, $h\neq h(T)$] the previous expression is drastically
simplified.

\vspace{0.2cm}
\noindent {\elevenit Bilinear scalar terms:}

It is not clear by now where a bilinear term
of the form $\mu H_1H_2$ in $W_{{\mathrm {obs}}}$ has its origin,
although it is well known that
this term is necessary in order to break $SU(2)\times U(1)_Y$
successfully. This is the so--called $\mu$ problem\footnote{Possible
solutions to the $\mu$ problem
can be found in refs.[18,9].}. Assuming that this term is actually present,
the corresponding bilinear term in the  scalar Lagrangian turns
out to be
\begin{eqnarray}
{\cal L}_{bil} & = & -m_{3/2}B \hat\mu H_1H_2 + \mathrm{h.c.} \nonumber \\
               & = & -m_{3/2}Y
 \left\{-\frac{\bar{W}_{S}}{\bar{W}}+
\left(3+
\frac{\bar{W}_{\bar S}}{\bar{W}} \frac{\delta^{GS}}{4
\pi^{2}} \right. \right.
                -  \left. \left.
\frac{\bar{W}_{\bar T}}{\bar{W}}
(T+\bar{T})\right) \right.
\nonumber \\
               & \times & \frac{1}{3Y+\frac{\delta^{GS}}{4\pi^{2}}}\left.
\left[ 3+n_{1}+n_{2}
- (T+\bar{T})
\frac{\mu_{T}}{\mu}\right]\right\}
\hat\mu H_{1}H_{2}\; +\; \mathrm{h.c.}
\label{Lbil}
\end{eqnarray}
with a notation similar to that of Eq.(\ref{Ltri}). Again,
$\hat \mu =
e^{K/2}\prod_{\alpha=1}^2 (K_{H_\alpha \bar H_\alpha})^{-1/2}
\mu$ is the
effective parameter giving, for example, the coupling between
the two higgsinos. For simplicity in the notation, we will drop
the hat from here on. Because of the above
mentioned ignorance about the origin of $\mu$, we prefer in the
following to consider $B$ as an unknown parameter.

The main conclusion at this stage is that some of the
common assumptions of the minimal supersymmetric standard model,
in particular universality for all the gaugino masses, scalar masses
and trilinear terms, do not hold in general (this
was already noted in ref.[17]). For $m^2_{{\phi}_{\alpha}}$
this lack of universality is related to the different values of
the modular weights ($n_\alpha$). Since large classes of fields
(e.g. for the $Z_3$ orbifold all the untwisted or all the twisted fields)
share the same modular weights, this is not necessarily dangerous
for flavor changing neutral currents.
Nevertheless, in order to get more quantitative results we need to know
$m_{3/2}$, $\langle W\rangle$, $\langle W_S\rangle$, etc.
This can only be done in the framework of a SUSY breaking scenario.
We turn to this point in the next section.
\vglue 0.6cm
{\elevenbf\noindent 3. Realization of SUSY breaking by gaugino condensation}
\vglue 0.4cm
As was mentioned above, the only mechanism so far analysed,
capable of generating a hierarchical SUSY breakdown in superstring
theories, is gaugino condensation in the hidden sector.
An extensive study of the status and properties of this mechanism can be
found in ref.[7]. Let us briefly summarize here the main
characteristics. For a hidden sector gauge group $G=\prod_bG_b$,
gaugino condensation induces a non--perturbative superpotential
\begin{eqnarray}
W^{np}=\sum_b d_b\ \frac{e^{-3k_bS/2\beta_b}}{[\eta(T)]^{6-
(3k_b\delta^{GS}/4\pi^2\beta_b)}}
\;\;,
\label{Wcond}
\end{eqnarray}
where $\beta_b$ are the corresponding beta functions and $d_b$ are
constants. Notice that $W^{np}_T=-2\eta^{-1}\frac{\partial \eta}{\partial T}
(3W^{np}+\frac{\delta^{GS}}{4\pi^2}W^{np}_S)$, which can be used to eliminate
$W_T$ in (\ref{M12}--\ref{Lbil}).
There is a large class of scenarios for which SUSY is spontaneously
broken (essentially through a non-vanishing $T$ F-term),
yielding reasonable values of $m_{3/2}$ and $\langle Y \rangle$
without the need of fine tuning${}^{7}$. A common ingredient of these
models is the existence of more than one condensing group
in the hidden sector. The values of $m_{3/2}$ and
$\langle Y \rangle$ turn out to depend almost exclusively on
which the gauge group and matter content of the hidden sector are.
Practically, any values of $m_{3/2}$ and $\langle Y \rangle$ are available by
appropriately choosing the hidden sector. Consequently, we can consider
$m_{3/2}$ and $Y$ as free parameters since no dynamical mechanism
is already known to select a particular string vacuum. On the
other hand, both $m_{3/2}$ and $\langle Y \rangle$ are almost independent
of the value of $\delta^{GS}$. Likewise, the value of
$\langle T \rangle$ turns out to be $\sim 1.23$ in all cases${}^{3,7}$.
This stability does not hold, however, for the other
quantities ($\langle W \rangle$, $\langle W_S \rangle$, etc.) appearing in
Eqs.(\ref{M12}--\ref{Lbil}). For example, as was pointed out in
ref.[3], the combination $YW_S-W$ (which is proportional to the
$S$ F-term and appears in several places in
the previous equations) is vanishing in the minimum of the
scalar potential for $\delta^{GS}=0$, but this is no longer true for
$\delta^{GS}\neq 0$. In this case, however, the scalar potential
is much more involved, so a numerical analysis is in general
necessary.
Fortunately, the results admit quite simple and useful
parameterizations describing them very well (within $1\%$ of accuracy).
Next, we give these parameterizations for the common $k^a=1$ case.

\vspace{0.2cm}
\noindent {\elevenit Scalar masses:}
\begin{eqnarray}
m_{\phi_\alpha}^2= m_{3/2}^2\left[1+n_\alpha(0.078-1.3\times 10^{-4}
\ \delta^{GS})\right]
\;\;.
\label{mapr}
\end{eqnarray}

\vspace{0.2cm}
\noindent {\elevenit Gaugino masses:}

For the $Z_3$ and $Z_7$ orbifolds the threshold contributions
to the $f$ function [see Eq.(\ref{fw})] are known to vanish${}^{12}$ and the
following equality holds for all the gauge group factors
\begin{eqnarray}
\sum_{i} b_i^{'a}\ -\ 2k^a\delta^{GS}=0
\;\;.
\label{cancel}
\end{eqnarray}
As a consequence, there is a cancellation of the second term of
Eq.(\ref{M12}). The value of the first term at the minimum of
the potential is easily parameterizable as
\begin{eqnarray}
M_a(Q) = -\alpha_{a}(Q)\ m_{3/2}\ (0.0120\ \delta^{GS}+0.019)
\;\;.
\label{Mapr}
\end{eqnarray}
Hence, for the $Z_3$ and $Z_7$ orbifolds the value of $M_a$
is completely given in terms of $m_{3/2}$ and $\delta^{GS}$. For
the rest of the $Z_N$ orbifolds things are different since Eq.(\ref{cancel})
is no longer true, allowing also for a lack of universality of
the gaugino masses. Then, an equation similar to Eq.(\ref{Mapr})
can be written in terms of $m_{3/2}$, $\delta^{GS}$ and
$\sum_i b_{i}^{'a}$. Let us note, however, that for all the remaining
$Z_N$ orbifolds (except for the $Z_6$--II) the cancellation (\ref{cancel})
still takes place for two of the three compactified dimensions\footnote{
For the $Z_6$--II case the cancellation holds in one complex dimension.
On the other hand, the $Z_6$--II orbifold is one of the less interesting
orbifolds from the phenomenological point of view${}^{19}$.}, i.e.
$
b_i^{'a}\ -\ 2k^a\delta_i^{GS}=0\;\;(i=1,2)
$.
Roughly speaking, our numerical results indicate that
parameterization (\ref{Mapr}) is still valid in these cases within
a $30\%$ error. As will be seen in the next section,
this is enough for our purposes.

\vspace{0.2cm}
\noindent {\elevenit Trilinear and bilinear scalar terms:}

If the three fields under consideration are untwisted, i.e.
$n_1=n_2=n_3=-1$, $A$ turns out to be very small in all the cases:
$
A^{untw}\stackrel{<}{{}_\sim} 10^{-3}<<1
$
(if $\delta^{GS}=0$, $A^{untw}=0$ exactly).
For twisted fields a good parameterization is
\begin{eqnarray}
A = A^{untw} + (0.28-2.3\times10^{-4} \delta^{GS})\left(3+\sum_{\alpha=1}^3
n_{\alpha}\right) + (0.69-6.9\times10^{-4} \delta^{GS}) \frac{h_{T}}{h}
\;\;,
\label{Atot}
\end{eqnarray}
where the precise value of $h_{T}/h$ depends on the specific
Yukawa coupling considered (see ref.[15]). In general $h_{T}/h$ is
negative and $|h_{T}/h|\stackrel{<}{{}_\sim}O(1)$.
Finally, as explained above, we choose
to leave the value of the bilinear term coefficient, $B$,
free owing to our ignorance of the origin of $\mu$ [see Eq.(\ref{Lbil})].

Equations (\ref{mapr},\ref{Mapr},\ref{Atot})
summarize the input of soft breaking terms to
be studied from the phenomenological point of view. Before entering
in a more detailed analysis, let us point out their most outstanding
characteristics. First the scalar masses are not universal (for fields
with different modular weights $n_{\alpha}$), but the deviation
from universality is very small (since $n_{\alpha}=O(1)$), essentially
consistent with flavor changing neutral currents${}^{20}$. Second,
gaugino masses tend to be much smaller than scalar masses. In fact,
for $\delta^{GS}=0$ the first term of Eq.(\ref{M12}) (proportional to the
$S$ F-term) vanishes at the minimum of the potential${}^{3}$. Then,
the resulting gaugino mass is very tiny, as is reflected
in Eq.(\ref{Mapr}), since it is generated thanks to the one-loop
threshold correction of Eq.(\ref{fw}). Actually, $\delta^{GS}\neq 0$
is itself a one-loop efect, so it is not surprising that $M_a<<m_\phi$.
Since typically${}^{13}$ $\delta^{GS}\stackrel{<}{{}_\sim} 50$,
this means $M_a\stackrel{<}{{}_\sim}\frac{1}{10}m_\phi$. This is probably
a quite general fact\footnote{Work in progress.} for a wide class of
non-perturbative superpotentials.
\vglue 0.6cm
{\elevenbf\noindent 4. Phenomenological viability of the soft breaking terms}
\vglue 0.4cm
{\elevenit\noindent 4.1. Experimental constraints}
\vglue 0.1cm
\baselineskip=14pt
\elevenrm

As was mentioned in the introduction, there are two types of
constraints on soft breaking terms: observational bounds and naturalness
bounds. The first ones mainly come from direct production of
supersymmetric particles in accelerators. This gives the following
lower bounds on supersymmetric particle masses, as reported by the
Particle Data Group${}^{21}$
\begin{eqnarray}
M_3 &>&79\ \mathrm{GeV}\;\;,\;\;\; M_{\chi^{\pm}}{>}45\ \mathrm{GeV}
\nonumber \\
m_{\tilde q}&>&74\ \mathrm{GeV}\;\;,\;\;\; m_{\tilde l}
\stackrel{>}{{}_\sim} 45\ \mathrm{GeV}
\;\;,
\label{mexp}
\end{eqnarray}
where $M_3$ is the gluino mass, $\chi^{\pm}$ is the lightest
chargino, and $\tilde q, \tilde l$ collectively denote squarks and sleptons
respectively\footnote{Less conservative bounds have been reported
elsewhere${}^{22}$. The corresponding modification of our final results
is completely straightforward.}.
Since in the stringy schemes we are analyzing gaugino masses are much
smaller than scalar masses, the relevant experimental (lower) bounds for
us are those corresponding to $M_3$ and $M_{\chi^{\pm}}$. In particular,
for the gluino mass [(see Eq.(\ref{Mapr})]
\begin{eqnarray}
M_3 \simeq \alpha_{3}(M_Z) m_{3/2}\ (0.0120\ \delta^{GS}+0.019)>79\
\mathrm{GeV}\;\;.
\label{M3apr}
\end{eqnarray}
This translates into lower bounds on $m_{3/2}$ and (using Eq.(\ref{mapr}))
$m_{\tilde q}$, $m_{\tilde l}$:
\begin{eqnarray}
m_{3/2} & \stackrel{>}{{}_\sim} & \frac{79\ \mathrm{GeV}}
{\alpha_{3}(M_Z)\ (0.0120\ \delta^{GS}+0.019)}
\nonumber \\
m_{\tilde{q}}(M_{Str}), m_{\tilde{l}}(M_{Str})
& \stackrel{>}{{}_\sim} & (79\ \mathrm{GeV})
\frac{\sqrt{1+n_{\alpha}(0.078-1.3\times10^{-4}
\delta^{GS})}}{\alpha_{3}(M_{Z})(0.0120 \delta^{GS}+0.019)}
\;\;,
\label{bound1}
\end{eqnarray}
Similar bounds, but
involving also $\mu$ and $\sin 2\beta$, are obtained from the lightest
chargino mass. However, these also involve the values of $\mu$
(whose origin is not clear yet) and $\sin 2\beta$ (where
${\mathrm{tg}} \beta=\langle H_2\rangle/\langle H_1\rangle$) and, hence,
they are not as clean as those of Eq.(\ref{bound1}).
\vglue 0.4cm
{\elevenit\noindent 4.2. Naturalness constraints}
\vglue 0.1cm
\baselineskip=14pt
\elevenrm

One of the most interesting features of low-energy supergravity is that
the electroweak breaking can arise as a direct consequence of SUSY
breaking${}^{23}$. More precisely, the corresponding scalar potential
for the Higgs fields is given by
\begin{eqnarray}
V(H_1,H_2)=\frac{1}{8}(g^2+g'^2)\left(|H_1|^2-|H_2|^2\right)^2
+ \mu_1^2|H_1|^2 + \mu_2^2|H_2|^2 -\mu_3^2(H_1H_2+\mathrm{h.c.})
\;\;,
\label{Vhiggs}
\end{eqnarray}
where $\mu_1$, $\mu_2$, $\mu_3$ are related to the soft breaking
parameters via well--known${}^{23}$ expressions. The minimization of
$V(H_1,H_2)$ yields $\langle H_1\rangle$, $\langle H_1\rangle$
and thus $M_Z(a_i,h_t)$, where $a_i$ are the theoretical parameters
defining the SUSY breaking and $h_t$ is the top Yukawa coupling, which
is taken as an independent parameter. For this process to be natural
one must impose the absence of fine tuning on the values of the
$a_i$ parameters. This is usually acomplished through the standard
criterion${}^{24}$
\begin{eqnarray}
\left|\frac{a_i}{M_Z^2}\frac{\partial M_Z^2(a_i,h_t)}
{\partial a_i}\right|<\Delta
\;\;,
\label{natur}
\end{eqnarray}
where $\Delta$ measures the allowed degree of fine tuning\footnote{
A discussion on the validity of this criterion can be found in ref.[25]}.
In ref.[24], in the context of the minimal supersymmetric standard
model (MSSM), the soft breaking parameters $M, m, \mu, A, B $ were
taken as the $a_i$ parameters of Eq.(\ref{natur}). In fact, in the MSSM,
the former are quantities whose origin is unknown rather than independent
parameters. On the contrary, in our stringy context assuming gaugino
condensation, these parameters (except $\mu$ and $B$) are related each
other via Eqs.(\ref{mapr}--\ref{Atot}). Consequently, in our case
the role of the independent $a_i$ parameters in (\ref{natur})
has to be played by $m_{3/2}$, $\mu$ and $B$. After the corresponding
analysis (for details see ref.[8]) we arrive from (\ref{natur})
to upper bounds on the the values of $m_{3/2}$, $\mu$ and $B$, and
thus on all the supersymmetric masses. Of course these bounds depend
on the values of $\Delta$ and $\delta^{GS}$. (Here we take
conservatively${}^{25}$ $\Delta=50$. On the other hand,
typically${}^{13}$
$\delta^{GS}\stackrel{<}{{}_\sim} 50$.) Besides these upper
bounds there are the experimental lower bounds obtained in the previous
subsection. This gives the allowed ranges of variation for all the
masses. For example, for a typical case ($\delta^{GS}=45$)
we find
\begin{eqnarray}
1100\ \mathrm{GeV}\leq m_{3/2}\leq 4200\ \mathrm{GeV} \;& , & \;\;
350\ \mathrm{GeV}\leq \mu
\leq 450\ \mathrm{GeV}
\nonumber \\
79\ \mathrm{GeV}\leq M_3\leq 285\ \mathrm{GeV} \;& , & \;\; 45 \
\mathrm{GeV}\leq
M_{\chi^{\pm}}\leq 80\ \mathrm{GeV}
\nonumber \\
110\ \mathrm{GeV}\leq m_{top}\leq 165\ \mathrm{GeV}\;& , & \;\;
1070\ \mathrm{GeV}\leq
m_{\tilde q}, m_{\tilde l}\leq 4080 \ \mathrm{GeV}
\;\;,
\label{cotas}
\end{eqnarray}
(remarkably, the above upper bound on $m_{top}$ arises naturally, without
using the current theoretical upper bound). As a general fact
we can say that the gluino and chargino masses
should be quite close to their present experimental limits, whereas
$m_{\tilde q}, m_{\tilde l}$ should be much higher
($\stackrel{>}{{}_\sim}\ 1$ TeV).
\vglue 0.6cm
{\elevenbf\noindent 5. Conclusions}
\vglue 0.4cm
The general form of the SUSY soft breaking terms is derivable
for a large class of phenomenologically interesting string
constructions, more precisely symmetric orbifolds.
As a general characteristic the soft breaking terms show
a certain lack of universality (unlike
the usual assumptions of the minimal supersymmetric standard model),
though this is not necessarily dangerous
for flavor changing neutral currents.
To get more quantitative results a specific SUSY breaking
mechanism has to be considered,
gaugino condensation being the most attractive on so far proposed.
The most striking property of the soft breaking terms
derived from gaugino condensation is that squark and slepton masses
tend to be much larger than gaugino masses
($m_{\phi}\stackrel{>}{{}_\sim}10M_a$), which probably is a quite general
fact. Experimental
bounds and the requirement of a successful electroweak breaking
without fine tuning impose further restrictions on the soft breaking
terms. As a consequence
the gluino and chargino masses
should
be quite close to their present experimental limits, whereas
squark and slepton masses
should be much higher
($\stackrel{>}{{}_\sim}\ 1$ TeV).
\vglue 0.5cm
{\elevenbf\noindent 7. References \hfil}
\vglue 0.4cm

%
%
%

\end{document}